\documentclass[conference]{IEEEtran}
\IEEEoverridecommandlockouts

\usepackage[table]{xcolor}
\usepackage{float} 

\usepackage{cite}
\usepackage{amsmath,amssymb,amsfonts}
\usepackage{algorithmic}
\usepackage{graphicx}
\usepackage{textcomp}
\usepackage{xcolor}
\usepackage{wasysym}

\usepackage{makecell, array}

\usepackage[ruled,vlined]{algorithm2e}

\usepackage{soul}
\usepackage{adjustbox}

\usepackage{hyperref}

\usepackage{helvet}

\usepackage{trimclip}

\usepackage{stackengine,scalerel}

\usepackage{array}        
\usepackage[utf8]{inputenc}  
\usepackage[greek,english]{babel}  

\usepackage{tcolorbox}

\usepackage{tgheros} 
\usepackage{tikz}

\definecolor{CPU}{rgb}{0.8,0,0}    
\definecolor{GPU}{rgb}{0,0,0.8}    
\definecolor{MPI}{rgb}{0,0.5,0}    

\def\BibTeX{{\rm B\kern-.05em{\sc i\kern-.025em b}\kern-.08em
    T\kern-.1667em\lower.7ex\hbox{E}\kern-.125emX}}
\begin{document}


\title{Exascale  Implicit Kinetic Plasma Simulations on El~Capitan for Solving the Micro-Macro Coupling in Magnetospheric Physics}

\author{\IEEEauthorblockN{Stefano~Markidis\textsuperscript{*}, Andong~Hu\textsuperscript{*}, Ivy~Peng\textsuperscript{*}, Luca~Pennati\textsuperscript{*},  Ian~Lumsden\textsuperscript{\textdagger}, \\ Dewi~Yokelson\textsuperscript{\S}, Stephanie~Brink\textsuperscript{\S}, Olga~Pearce\textsuperscript{\S}, Thomas~R.W.~Scogland\textsuperscript{\S}, \\ Bronis~R.~de~Supinski\textsuperscript{**}, Gian Luca Delzanno\textsuperscript{\ddag}, Michela~Taufer\textsuperscript{\textdagger}}
\IEEEauthorblockA{
\textsuperscript{*}School of Electrical Engineering and Computer Science, KTH Royal Institute of Technology, Stockholm, Sweden \\
\textsuperscript{\textdagger} Department of Electrical Engineering and Computer Science, University of Tennessee, Knoxville, TN, USA\\
\textsuperscript{\S}Center for Applied Scientific Computing, Lawrence Livermore National Laboratory, Livermore, CA, USA \\
\textsuperscript{**}Livermore Computing, Lawrence Livermore National Laboratory, Livermore, CA, USA \\
\textsuperscript{\ddag} T-5 Applied Mathematics and Plasma Physics, Los Alamos National Laboratory, Los Alamos, NM, USA}

}

\maketitle

\begin{abstract}
Our fully kinetic, implicit Particle-in-Cell~(PIC) simulations of global magnetospheres
on up to 32,768 of El Capitan's AMD Instinct MI300A Accelerated Processing Units~(APUs) 
represent an unprecedented computational capability that addresses a fundamental challenge in space 
physics: resolving the multi-scale coupling between microscopic (electron-scale) and macroscopic 
(global-scale) dynamics in planetary magnetospheres. The implicit scheme of \texttt{{\small iPIC3D}}  
supports time steps and grid spacing up to 10× larger than explicit methods, without sacrificing
physical accuracy, enabling simulation of  magnetospheres while preserving fine-scale electron 
physics critical for key processes like magnetic reconnection and plasma turbulence. Our algorithmic 
and technological innovations include GPU-optimized kernels, particle-control, and physics-aware data 
compression using Gaussian Mixture Models. With simulation domains spanning 100–1,000 ion skin depths, 
we reach the global scale of small-to-medium planetary magnetospheres, such as those of Mercury, 
and Ganymede, which supports full-kinetic treatment of global-scale dynamics in systems previously out 
of reach for fully kinetic PIC codes.
\end{abstract} 


\section{Introduction}
A magnetosphere is the highly dynamic space environment around a planet dominated by its magnetic field and shaped by its interaction with a plasma wind~\cite{russell1972configuration,gold1959motions}. Earth's magnetosphere plays an important role in shielding space- and ground-based technologies from potentially catastrophic space weather effects. Disturbances such as magnetic storms can disrupt spacecraft operations, communication systems, and terrestrial infrastructure, including power grids and pipelines~\cite{baker1998space}. For these reasons, modeling magnetospheres is important not only for advancing fundamental plasma physics but also for developing predictive capabilities to mitigate space weather impacts~\cite{lapenta2012space,innocenti2017progress}.
\begin{figure}[t]
    \centering
    \includegraphics[width=0.95\linewidth]{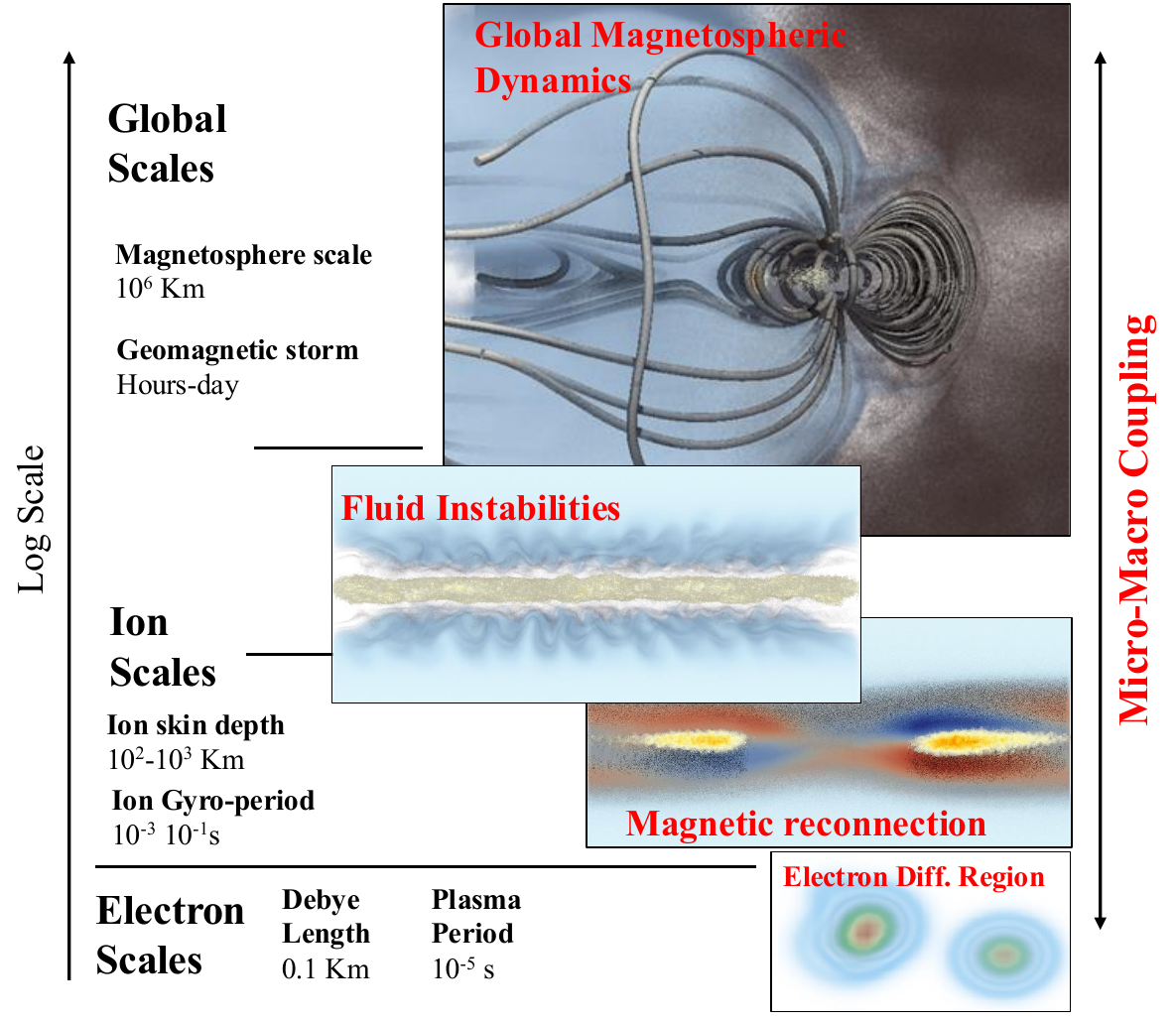}  
    \caption{Different spatial and temporal scales in the Earth's magnetosphere. Localized electron-scale phenomena can couple to and drive global-scale magnetospheric dynamics.}
    \label{fig:micromacro}
\end{figure}

Despite decades of research, the first-principles simulation of large planetary magnetosphere, such as Earth's, in realistic physical parameters remains a grand challenge in computational plasma physics. Such simulations present two fundamental challenges. First, the disparity in mass between electrons and ions (positively charged particles, such as protons) -- approximately a factor of 1,836 -- leads to largely different spatial and temporal scales. Ions govern global-scale dynamics, while electrons dominate localized, high-frequency processes. Second, energy and momentum are exchanged continuously across these scales. This \textit{micro-macro coupling} implies that micro-scale processes can influence, or be influenced by, the global magnetospheric state~\cite{lapenta2006kinetic,wan2008micro,lapenta2013swiff}. An example is magnetic reconnection~\cite{biskamp1996magnetic} -- a fundamental plasma process in which magnetic field lines break and reconnect, rapidly converting magnetic energy into particle acceleration, plasma heating, and bulk flows. Though magnetic reconnection originates in narrow electron-scale regions (known as \textit{electron diffusion layers}), reconnection can trigger large-scale topological changes in the entire magnetosphere. Conversely, large-scale reconfigurations of the magnetosphere, such as fluid instabilities, can drive secondary magnetic reconnection events in localized regions~\cite{lapenta2015secondary}. 

At the largest scales, the magnetosphere structure is shaped by the fluid balance between planetary magnetic pressure and the incident plasma flow, resulting in a characteristic comet-like shape: compressed on the dayside (the body side directly exposed to the wind), elongated into a magnetotail on the nightside~\cite{borovsky2018earth,spreiter1966hydromagnetic}. However, magnetospheric plasmas are far from ordinary fluids. They consist of charged particles tightly coupled to electromagnetic fields and capable of collective behavior. As a result, localized kinetic processes such as wave–particle interactions, plasma turbulence~\cite{olshevsky2013energetics,olshevsky2016magnetic} and reconnection propagate across scales, shaping global dynamics. 

Quantifying the spatial extent of planetary magnetospheres is essential for comparing systems of different scales and guiding the design of global magnetospheric simulations. To facilitate comparative studies across planetary systems, magnetospheric sizes are often expressed in terms of the local solar wind ion skin depth, $d_i$, a characteristic scale that reflects ion inertial effects in plasmas and is approximately 100 km. Two key measures are typically used: the \textit{standoff distance} -- the distance from the planetary center to the subsolar point where solar wind pressure balances the planetary magnetic field -- and the \textit{magnetopause size}, which defines the overall extent of the magnetospheric boundary. Although usually measured in kilometers, expressing these distances in units of $d_i$ provides a normalized framework for comparing different planetary magnetospheres. Table~\ref{tab:magnetosphere_sizes} summarizes representative magnetopause sizes across the solar system in this form. It is important to note that these distances reflect the extent of the magnetopause boundary alone; the whole computational domain required to capture magnetospheric dynamics realistically is typically at least four to five times larger in each dimension.  Small and medium systems, such as those of Mercury and Ganymede, have magnetopause sizes ranging from approximately 10 to 50 $d_i$. Earth's magnetosphere extends beyond 600 $d_i$.  Modeling even these compact systems in a fully kinetic, three-dimensional framework with realistic parameters has historically been computationally prohibitive.

\begin{table}[t]
\centering
\caption{Approximate magnetopause sizes for various planetary bodies, expressed in ion skin depths (\( d_i \)), assuming \( d_i \approx 100 \, \text{km} \).}
\renewcommand{\arraystretch}{1.3}
\begin{adjustbox}{max width=\columnwidth,center}
\large
\begin{tabular}{lccc}
\hline
\textbf{Body} & \textbf{Standoff Distance (km)} & \textbf{Size in \( d_i \)} & \textbf{Notes} \\
\hline
\rowcolor{gray!20}
Mercury        & ~1,500                  & ~15          & Intrinsic magnetosphere \\
Ganymede 
& ~2,000–2,500            & ~20–25       & Intrinsic magnetosphere \\
\rowcolor{gray!20}
Mars           & ~1,000–2,000            & ~10–20       & Induced magnetosphere \\
Earth          & ~64,000    & ~640         & Large intrinsic magnetosphere \\
\rowcolor{gray!20}
Jupiter        & \( > 3 \times 10^6 \)  & \( >5,000 \) & Giant system \\
\hline
\end{tabular}
\end{adjustbox}
\label{tab:magnetosphere_sizes}
\end{table}

This paper targets magnetosphere simulation domains spanning 100–1,000 ion skin depths, reaching the global scale of small-to-medium planetary magnetospheres, such as those of Mercury and Ganymede. Enabled by algorithmic and architectural innovations, our simulations resolve both large-scale structure and kinetic microphysics self-consistently, using the \texttt{{\small iPIC3D}} implicit Particle-In-Cell (PIC) code. We scale \texttt{{\small iPIC3D}} to 32,768 AMD Instinct MI300A APUs on El Capitan, demonstrating an end-to-end simulation capability that includes implicit time integration, hybrid CPU–GPU workflows, dynamic particle control, and integrated data analysis. Taken together, these developments enable a new modeling regime in space plasma physics—one in which small-to-medium planetary magnetospheres can be simulated from first principles at scale, with realistic parameters.



\section{Related Work}
Multiple plasma simulation approaches have been developed to model magnetospheric dynamics, spanning from fluid-based models to fully kinetic PIC methods. Fluid-based models--such as magnetohydrodynamics (MHD) and extended MHD--remain the primary workhorses for global-scale simulations. Established frameworks, such as the Space Weather Modeling Framework (SWMF)~\cite{toth2005space,gombosi2021sustained}, implement these models and some extensions. However, fluid models inherently cannot resolve key kinetic phenomena, including wave–particle interactions, plasma turbulence, and magnetic reconnection. For example, magnetic reconnection in the collisionless regime is not physically supported in ideal MHD due to the frozen-in condition, which prohibits field line breaking. In practice, reconnection is often observed in such models through numerical artifacts, such as artificial resistivity. To capture these processes accurately, simulation approaches must retain the coupling between microscopic (electron-scale) and macroscopic (magnetosphere-scale) physics—that is, the micro–macro coupling.

\begin{table}[t]
    \centering
    \caption{Advanced capabilities for magnetospheric kinetic HPC simulation codes. The \LEFTcircle$~$symbol indicates partial capability.}
    \renewcommand{\arraystretch}{1.3}
    \begin{tabular}{lccccccc}
        \hline
        Capability & \rotatebox{90}{Tristan} & \rotatebox{90}{H-VPIC} & \rotatebox{90}{AMITIS} & \rotatebox{90}{A.I.K.E.F.} & \rotatebox{90}{Vlasiator} & \rotatebox{90}{MHD-EPIC} & \rotatebox{90}{iPIC3D} \\
        \hline
        Fully Kinetic & \checkmark & \LEFTcircle & \LEFTcircle & \LEFTcircle & \LEFTcircle & \LEFTcircle & \checkmark \\
        Implicit Discretization &  &  & \checkmark & \checkmark & \checkmark & \LEFTcircle & \checkmark \\
        Dyn. LB / Particle Control & \checkmark  &  &  & \checkmark & \checkmark & \LEFTcircle & \checkmark \\
        GPU Acceleration &  & \checkmark & \checkmark & & \LEFTcircle & \LEFTcircle & \checkmark \\
        In-Situ Compression &  &  &  &  &  &  & \checkmark \\
        Integrated Data Analysis & \checkmark & \LEFTcircle &  &  & \LEFTcircle &  & \checkmark   \\
        \hline
    \end{tabular}
    \label{tab:magneto_code_capabilities}
\end{table}

Fully kinetic approaches, which solve the Vlasov–Maxwell system directly via PIC methods, offer this level of physical fidelity. Historically, these methods for magnetospheric simulations have been used by codes such as \texttt{{\small Tristan}}~\cite{nishikawa2001tristan,buneman1992solar,baraka2007sensitivity,Spitkovsky2019tristanMP}, initially developed by Buneman at NCSA. However, these explicit PIC codes were subject to stringent stability constraints, requiring resolution of the Debye length and plasma frequency, which in turn necessitates artificially large thermal velocities, large grid spacings, and exceedingly small time steps.

One widely adopted compromise is to relax the fully kinetic description by treating electrons as a fluid while retaining ion kinetics. These models, known as hybrid approaches, eliminate numerical instabilities associated with electron timescales and are more computationally tractable. Prominent hybrid codes include \texttt{{\small Vlasiator}}~\cite{von2014vlasiator}, \texttt{{\small Hybrid-VPIC}}~\cite{le2023hybrid}, \texttt{{\small AMITIS}}~\cite{fatemi2017amitis}, and \texttt{{\small A.I.K.E.F.}}~\cite{muller2011aikef}. While successful in capturing ion-scale dynamics over large domains, hybrid methods are not fully self-consistent and do not describe electron-scale physics.
A second strategy to balance scale fidelity and computational feasibility is the two-way coupling of global MHD models with localized kinetic regions. This technique, known as MHD with Embedded PIC (MHD-EPIC)~\cite{daldorff2014two,Yinsi2021MHDAEPIC}, embeds fully kinetic PIC simulations within extended MHD domains, targeting specific regions where kinetic effects are known to be important. MHD-EPIC has been used to study the magnetospheres of Ganymede~\cite{toth2016extended,zhou2019embedded}, Mercury~\cite{chen2019studying}, and Earth~\cite{chen2020magnetohydrodynamic}. While powerful, this approach faces some challenges. First, it requires dynamically identifying kinetic regions via predefined monitor functions and spawning new PIC instances~\cite{Yinsi2021MHDAEPIC}. Second, it assumes that the fluid-kinetic boundary can be described using Gaussian distribution functions, which may not capture complex particle distributions. Third, overlapping fluid and kinetic regions must be consistently coupled to maintain physical and numerical consistency~\cite{markidis2014fluid}.

Table~\ref{tab:magneto_code_capabilities} summarizes the capabilities of selected codes used for magnetospheric modeling, with emphasis on numerical discretization, dynamic load balancing, GPU acceleration, in-situ data compression, and integrated analysis. The \texttt{iPIC3D} code used in this work distinguishes itself by supporting fully kinetic physics with implicit time integration, particle control, accelerator support, and integrated in-situ workflows—all critical enablers for exascale kinetic simulations.

\section{Methodology}

\subsection{Algorithmic Contribution I: Implicit Discretization for Large-Scale Kinetic Plasma Simulations.}
Simulating magnetospheric dynamics from first principles requires capturing both micro- and macro-scale physics in a fully kinetic framework. This demands solving the Vlasov equation, which governs the evolution of the phase-space distribution function $f(\mathbf{x}, \mathbf{v}, t)$ for electrons and ions in a collisionless plasma, coupled with Maxwell’s equations. Unlike fluid models, kinetic approaches are capable of resolving velocity-space effects critical for phenomena such as wave-particle interactions and magnetic reconnection.

PIC methods are widely adopted for kinetic plasma simulations~\cite{birdsall2018plasma,dawson1983particle}. They discretize $f$ into computational particles whose trajectories evolve under self-consistent electromagnetic fields. The relativistic equations of motion for each particle are:
\begin{equation}
    m_s \frac{d (\gamma\mathbf{v}_p)}{dt} = q_s \left( \mathbf{E}_p + \frac{\mathbf{v}_p \times \mathbf{B}_p}{c} \right), \quad \frac{d\mathbf{x}_p}{dt} = \mathbf{v}_p,
\end{equation}
where $m_s$ and $q_s$ are the particle mass and charge, $\mathbf{x}_p$ and $\mathbf{v}_p$ are particle position and velocity, $c$ is the speed of light, $\gamma~=~1/\sqrt{1-\mathbf{v}^2_p/c^2}$ is the Lorentz relativistic factor, and $\mathbf{E}_p$ and $\mathbf{B}_p$ are the interpolated electric and magnetic fields at the particle location. Fields are solved on a spatial grid using interpolated charge and current densities from particles. Explicit PIC methods solve these equations using FDTD schemes and the Boris mover~\cite{birdsall2018plasma}, but are limited by stability constraints (Debye length, plasma frequency), which restrict the timestep and grid spacing. To overcome these limitations, we employ an implicit discretization strategy that allows for larger time steps and coarser grids while maintaining accuracy. 

\noindent \textbf{Implemented Solution:} Our code, \texttt{{\small iPIC3D}}~\cite{markidis2010multi}, employs a \textit{moment-implicit} PIC formulation. Written in C++, \texttt{{\small iPIC3D}} is optimized for HPC systems via domain decomposition on a 3D Cartesian grid. Inter-node communication is handled via MPI, while intra-node parallelism is achieved using OpenMP. The code is vectorized and designed for portability across CPU and GPU architectures, including AMD and NVIDIA accelerators, via different backends. In addition, the iPIC3D code includes advanced open boundary conditions, sub-cycling techniques for the particle movers, and relativistic extensions, which are not discussed in this paper.

\noindent \textit{Particle Mover:} The particle motion is advanced using a relativistic predictor-corrector algorithm. This method computes the average particle velocity using both electric and magnetic field contributions, and includes a guiding-center correction for gyro-motion~\cite{peng2015energetic}:
\begin{equation}
\left\{
\begin{alignedat}{1}
    &\mathbf{x}_{p}^{n+1} = \mathbf{x}_{p}^{n} + \bar{\mathbf{v}}_{p} \Delta t, \\
    &\mathbf{v}_p^{n+1} = \frac{\bar{\mathbf{v}}_p(\gamma^{n+1} + \gamma^n) - \gamma^n \mathbf{v}_p^n}{\gamma^{n+1}}, \\
    &\tilde{\mathbf{v}}_p = \gamma^n\mathbf{v}_p^n + \frac{q_s}{m_s}\frac{\Delta t }{2} \mathbf{E}_p^{n+1}, \\
    &\bar{\mathbf{v}}_p = 
   \underbrace{\frac{{\tilde{\mathbf{v}}_p}}{D}}_{\substack{\text{E Field Kick}}}
    + \underbrace{ \frac{\frac{\Delta t}{2\tilde{\gamma}}\, \tilde{\mathbf{v}}_{p} \times \boldsymbol{\Omega}^{n+1}_s}{D}}_{\substack{\text{Gyro-Rotation}}} 
     + \underbrace{\frac{\left(\frac{\Delta t}{2}\right)^2 \left(\frac{\tilde{\mathbf{v}}_{p}}{\tilde{\gamma}} \cdot \boldsymbol{\Omega}^{n+1}_s\right) \boldsymbol{\Omega}^{n+1}_s}
    {D}}_{\substack{\text{Guiding Center Correction}}}.
\end{alignedat}
\right.
\end{equation}
Here, $\bar{\mathbf{v}}_p$ is the average velocity, $\tilde{\mathbf{v}}_p$ is an intermediate velocity, $\mathbf{E}^{n+1}_p$ and $\boldsymbol{\Omega}^{n+1}_s = (q_s / cm_s) \mathbf{B}^{n+1}_p$ are electric and magnetic fields at particle $p$ position. The factors $\tilde{\gamma}$ and $D$ are respectively defined as $\tilde{\gamma}\approx (\gamma^n + \gamma^{n+1})/2$ and  
$D~=~\tilde{\gamma}[1 + ({\Delta t}/{(2\tilde{\gamma})} \boldsymbol{\Omega}^{n+1}_s)^2]$. The algorithm iteratively computes $\bar{\mathbf{v}}_p$ using fixed-point iteration. The \ul{compute-bound} particle mover kernel routine exhibits high data locality.

\noindent \textit{Implicit Maxwell Solver:} The electric and magnetic fields are updated using a moment-based formulation of Maxwell’s equations. Fluid moments—charge density $\rho$, current density $\mathbf{J}$, and the pressure tensor $\overline{\overline{\Pi}}$ —are computed by interpolating particle data to the grid:
\begin{equation}
\{ \rho, \mathbf{J}, \overline{\overline{\Pi}} \}_g = \sum_s \sum_p q_s \{1, \mathbf{v}_p, \mathbf{v}_p \mathbf{v}_p \} \mathcal{W}(\mathbf{x} - \mathbf{x}_p),
\end{equation}
where $\mathcal{W}$ is the interpolation function. The field update equation is:
\begin{multline}
\hspace{-1em}
\underbrace{%
 \underbrace{\left(\overline{\overline{I}}+ \overline{\overline{\chi}}^n\right) \cdot \mathbf{E}^{n+1}}_{\text{Implicit Medium Response}} 
  - (c\Delta t)^2 \underbrace{\Bigl(\nabla^2\mathbf{E}^{n+1} + \nabla\nabla\cdot\Bigl(\overline{\overline{\chi}}^n\cdot\mathbf{E}^{n+1}\Bigr)\Bigr)}_{\text{Wave Propagation and Diffraction}}%
}_{\text{Spectral Compression and Selective Damping}} \\
=\underbrace{\mathbf{E}^n + c\Delta t\Bigl(\nabla\times\mathbf{B}^n-\frac{4\pi}{c}\mathbf{\hat{J}}^n\Bigr)}_{\text{Faraday \& Ampere Law}}
-\underbrace{(c\Delta t)^2\nabla\Bigl(4\pi\hat{\rho}^n\Bigr)}_{\text{Poisson Law}}
\label{MaxwellSolver}
\end{multline}
where $\overline{\overline{\chi}}$ is the implicit dielectric tensor. $\overline{\overline{\chi}}$ is defined as:
\begin{equation}
\overline{\overline{\chi}}^n \cdot = \sum_{s} \overline{\overline{\chi}}^n_s \cdot,
\quad 
\overline{\overline{\chi}}^n_s \cdot \equiv 
\underbrace{\frac{1}{2} (\omega^n_{ps} \Delta t)^2}_{\text{ES Imp. Dielectric}} 
\underbrace{\overline{\overline{R_s}} \left(\mathbf{\Omega}^n_s \frac{\Delta t}{2}\right)}_{\text{Permeability Adj.}} \cdot
\end{equation}
where $\omega^n_{ps} = \sqrt{4 \pi \rho q_s/m_s }$ is the plasma frequency for the species $s$, $\mathbf{\Omega}_s^n \equiv q_s\mathbf{B}^n/(m_s c) $, and the tensor $\overline{\overline{R_s}}$ is defined as
$\overline{\overline{R_s}} = \overline{\overline{I}} - \frac{\Delta t}{2}\,[\mathbf{\Omega}_s]_\times + \left(\frac{\Delta t}{2}\right)^2\,\mathbf{\Omega}_s\,(\mathbf{\Omega}_s)^T$, with I as identity matrix and $[]_\times$ as cross–product matrix.

The corrected $\hat{\rho}, \hat{\mathbf{J}}$ densities are:
\begin{equation}
\hat{\rho}^n = \rho^n - \Delta t \nabla \cdot \hat{\mathbf{J}}^n, \quad \hat{\mathbf{J}}^n = \overline{\overline{R}} \cdot \left( \mathbf{J}^n - \frac{\Delta t}{2} \nabla \cdot \overline{\overline{\Pi}}^n \right).
\end{equation}
The field is computed with a flexible \ul{matrix-free GMRes solver}, improved with a variable preconditioner that uses few GMRes iterations~\cite{saad1993flexiGMres}. The linear solver is characterized by sparse computation, with irregular data accesses. Because of the ghost-cell exchange, network communication tends to dominate the Maxwell solver runtime, in case each domain is relatively small, e.g., strong scaling tests at very large node counts, making the \texttt{{\small iPIC3D}} Maxwell solver \ul{network-bound}.  The implicit PIC method is based on the \ul{spectral compression} and \ul{selective damping} of unresolved high-frequency modes, stabilizing the solution and preventing finite-grid instabilities~\cite{brackbill1982implicit,brackbill1985simulation}. After the electric field is computed, the new value of the magnetic field is advanced as:
\begin{equation}
\mathbf{B}^{n+1} = \mathbf{B}^n - c\Delta t \nabla \times \mathbf{E}^{n+1}.
\end{equation}

\noindent \textbf{Key Results and Impact:} The implicit formulation in \texttt{{\small iPIC3D}} enables time steps and grid spacings at least $10\times$ larger than those required by explicit methods \cite{markidis2010multi,lapenta2012space}. For 3D PIC simulations, this translates into a $10^4\times$ reduction in resolution requirements, retaining fidelity in electron-scale physics. 

\subsection{Algorithmic Contribution II: Dynamic and Selective Control of the Number of Particles} 

In magnetosphere PIC simulations, the domain boundaries are open.
From the inflow regions, wind electrons and protons are injected with a prescribed bulk velocity to convect the magnetic field~\cite{peng2015formation}, while the other boundaries serve as outflow regions where exiting particles are removed. 
The planet's dipole magnetic field, located at the center of the domain, acts as an obstacle, creating a bow shock in the case of a supersonic wind, which reflects particles and causes local accumulations. At the same time, in regions such as the magnetotails, the open BCs combined with particle drift decrease the particle count.
Consequently, some MPI processes handle up to twice or more the initial number of particles, while other MPI domains are greatly depleted. The resulting imbalance across subdomains is critical for computational performance, as uneven workloads force some MPI ranks to wait for others, wasting resources. Additionally, a low number of particles per cell results in a poor statistical representation of the distribution functions, which increases noise and affects the accuracy of wave-particle interactions, such as kinetic wave damping and instabilities~\cite{peng2015formation}.

\noindent \textbf{Implemented Solution:} We maintain accuracy with a balanced load by implementing a hybrid CPU-GPU particle control algorithm that monitors and adjusts the number of particles for each species in each MPI domain at runtime. At each cycle, we compare the current particle count with the initial number of particles in the domain ($N_p$), and when the deviation exceeds a threshold value $\theta$ (e.g., a 5\% change), we perform one of two operations following the method of Lapenta and Brackbill~\cite{lapenta1994dynamic}:
\begin{itemize}
    \item \textbf{Particle Coalescence -- \# particles $\boldsymbol{>}$ $\boldsymbol{N_p} (1 + \theta$):}  
    In cells with an excessive number of particles, we perform pair-wise merging between particles that are close in the phase space by combining their statistical weights. This operation reduces the particle count while preserving key physical properties.
    \item \textbf{Particle Splitting -- \# particles $\boldsymbol{<}$ $\boldsymbol{N_p} (1 - \theta$):} 
    In regions with an insufficient number of particles, we randomly select particles and split each into multiple particles, adjusting their statistical weights accordingly. Thus, we improve the resolution of the distribution function in underpopulated areas.
\end{itemize}
The coalescence operation requires particle sorting to identify particles that are close in phase space. Sorting in the position domain is performed on the CPU, while sorting in the velocity domain and the actual merging are handled by a GPU kernel. The splitting operation is entirely performed on the GPU. Unlike other particle control techniques~\cite{Faghihi2020constrainedResampling, chen2021unsupervised}, the method that we use~\cite{lapenta1994dynamic} does not require the solution of any linear system or quadratic problem and thus, is more suitable for an HPC runtime implementation.

\noindent \textbf{Key Results and Impact:}  The dynamic particle control scheme ensures balanced computational loads across MPI domains and improves the statistical accuracy of the distribution functions. The particle monitor is implemented as a local operation and does not require global reductions, thus adding little overhead. The coalescence operation is the most computationally expensive step in the algorithm, as it requires particle sorting. However, the hybrid CPU-GPU implementation allows us to overlap sorting with other operations (such as moving other species) and maximizes resource utilization. As a result, \ul{we achieve up to 10\% speed-up} (including the computational overhead of particle splitting and coalescence) for magnetic reconnection simulations, compared to simulations without control of the number of particles. 

\subsection{Technical Contribution I: GPU Acceleration and Hybrid Workflow}


The PIC method is well-suited for acceleration using GPUs due to its high degree of parallelism, particularly in the context of global magnetosphere simulations where large numbers of particles and grid points are involved. Key computational phases, such as the particle mover and particle-to-grid interpolation, are dominant in runtime and map efficiently onto accelerator architectures. These kernels are inherently parallel, with each particle independently updating its velocity and position or contributing to grid-based quantities. 

\noindent \textbf{Implemented Solution:} Efficient use of heterogeneous hardware is critical for large-scale magnetospheric simulations. Thus, we implement a hybrid workflow in \texttt{{\small iPIC3D}} (Algorithm~\ref{alg:ipic3d}), balancing the computational load between CPU and GPU components. The simulation is initialized with planetary magnetic dipole fields, a solar wind-convected magnetic field, and corresponding particle distributions in phase space. A typical simulation comprises $10^3$--$10^4$ time steps, with checkpointing and restart handled through the ADIOS 2 framework.

The main simulation loop is organized into three phases: \
\emph{(i) Phase 1: Particle Mover - Control \& I/O.} The predictor-corrector mover is executed on the GPU for each particle species in parallel using CUDA/HIP multi-streams, while I/O operations run concurrently on the CPU. Before the particle push, the hybrid CPU-GPU particle control algorithm adjusts the number of particles at the subdomain level for any species that require splitting or coalescence.
\emph{(ii) Phase 2: Interpolation \& MPI Exchange.} Interpolation of particle quantities ($\rho_s$, $J_s$, $\Pi_s$) occurs on the GPU. GPU-shared memory is used to minimize global memory traffic, and atomic operations ensure consistency. A first kernel processes particles that do not leave the local domain, while the exiting particles are transferred using MPI. Once the communication ends, a second kernel interpolates the transferred particles.
\emph{(iii) Phase 3: Maxwell Solver, Compression \& Data Analytics.} The CPU solves Maxwell's equations using a preconditioned GMRes solver, while the GPU performs in-situ data compression.

\begin{algorithm}
\DontPrintSemicolon
\caption{Accelerated \texttt{{\small iPIC3D}} Workflow}
\label{alg:ipic3d}
\SetAlgoLined
Initialize fields and particle data \tcp*[r]{\textbf{\textcolor{CPU}{(CPU)}}}\;
\For{each time step}{
  \SetKwBlock{phaseone}{Phase 1: Particle Mover - Control \& I/O}{}
  \phaseone{
    \begin{tabular}{>{\raggedright\arraybackslash}p{0.4\linewidth} >{\raggedleft\arraybackslash}p{0.4\linewidth}}
      \textbf{\textcolor{GPU}{GPU:}} &  \textbf{\textcolor{CPU}{CPU:}} \\
      Particle coalescence & Particle sorting \\
      Particle merging  \\
      \For{each species \(s\)}{Predictor-Corrector Mover} 
      &
      Parallel I/O
    \end{tabular}
  }
  
  \SetKwBlock{phasetwo}{Phase 2: Interpolation \& MPI Exchange}{}
  \phasetwo{
    \begin{tabular}{>{\raggedright\arraybackslash}p{0.45\linewidth} >{\raggedleft\arraybackslash}p{0.35\linewidth}}
      \textbf{\textcolor{GPU}{GPU:}}  & \textbf{\textcolor{MPI}{CPU/MPI:}}\\
      Flag exiting particles\\
      \For{each species \(s\)}{Interpolate (\(\rho_s\), \(J_s\), \(\Pi_s\)) stayed particles}
      &
      Transfer flagged particles \\
      \textbf{\textcolor{GPU}{GPU:}} & \\
    \multicolumn{2}{>{\raggedright\arraybackslash}p{\linewidth}}{
      Finalize interpolation received particles}
    \end{tabular}
  }
 
 \vspace{1em}
  \SetKwBlock{phasethree}{Phase 3: Maxwell Solver \& Compression \& Data Analysis}{}
  \phasethree{
    \begin{tabular}{>{\raggedright\arraybackslash}p{0.4\linewidth} >{\raggedleft\arraybackslash}p{0.4\linewidth}}
      \textbf{\textcolor{GPU}{GPU:}} & \textbf{\textcolor{MPI}{CPU/MPI:}} \\
      Compression  & Preconditioned GMRes  \\ 
      and GMM analysis  
    \end{tabular}
  }
}
\end{algorithm}
\begin{figure*}[t]
    \centering
    \includegraphics[width=\linewidth]{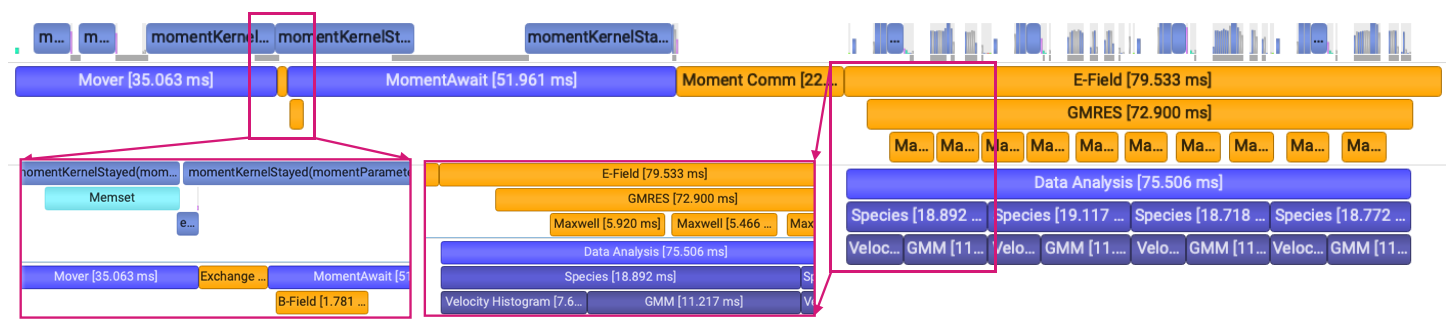}  
    \caption{Profiling of a single iPIC3D time step on an Nvidia Grace Hopper Superchip using the Nvidia Nsight tool. Blue labels indicate GPU kernels, while yellow labels indicate CPU kernels, demonstrating an almost perfect overlap of operations between the CPU and GPU, with high GPU utilization.}
    \label{fig:profiling}
\end{figure*}

The data layout is tuned to support different system architectures. In discrete host-device setups, particle and interpolated quantities reside in GPU memory, while field data remains in CPU memory and is transferred as needed. In unified memory systems, such as AMD MI300As, CPU and GPU capabilities share a common memory pool, eliminating redundant data transfers and reducing memory footprint. This sharing enables simulations with larger memory capacities, such as the 33-trillion-particle run performed on El Capitan.

The GPU backend of \texttt{{\small iPIC3D}} is implemented in CUDA. For AMD systems, it is translated to HIP via a header file that maps function calls. The code has been tested and optimized across multiple platforms, including AMD MI300A, AMD MI250x, NVIDIA Grace-Hopper Superchip, and Nvidia Ampere.

\noindent \textbf{Key Results and Impact:} Our optimizations attempt to \ul{maximize GPU/APU utilization} across heterogeneous architectures. Figure~\ref{fig:profiling} shows a detailed profile of a simulation step on a Grace Hopper Superchip. GPU kernels (blue) and CPU kernels (yellow) operate in parallel, with overlapping execution of the Maxwell solver on the CPU and in-situ analytics on the GPU.

\begin{figure}[t]
    \centering
    \includegraphics[width=\linewidth]{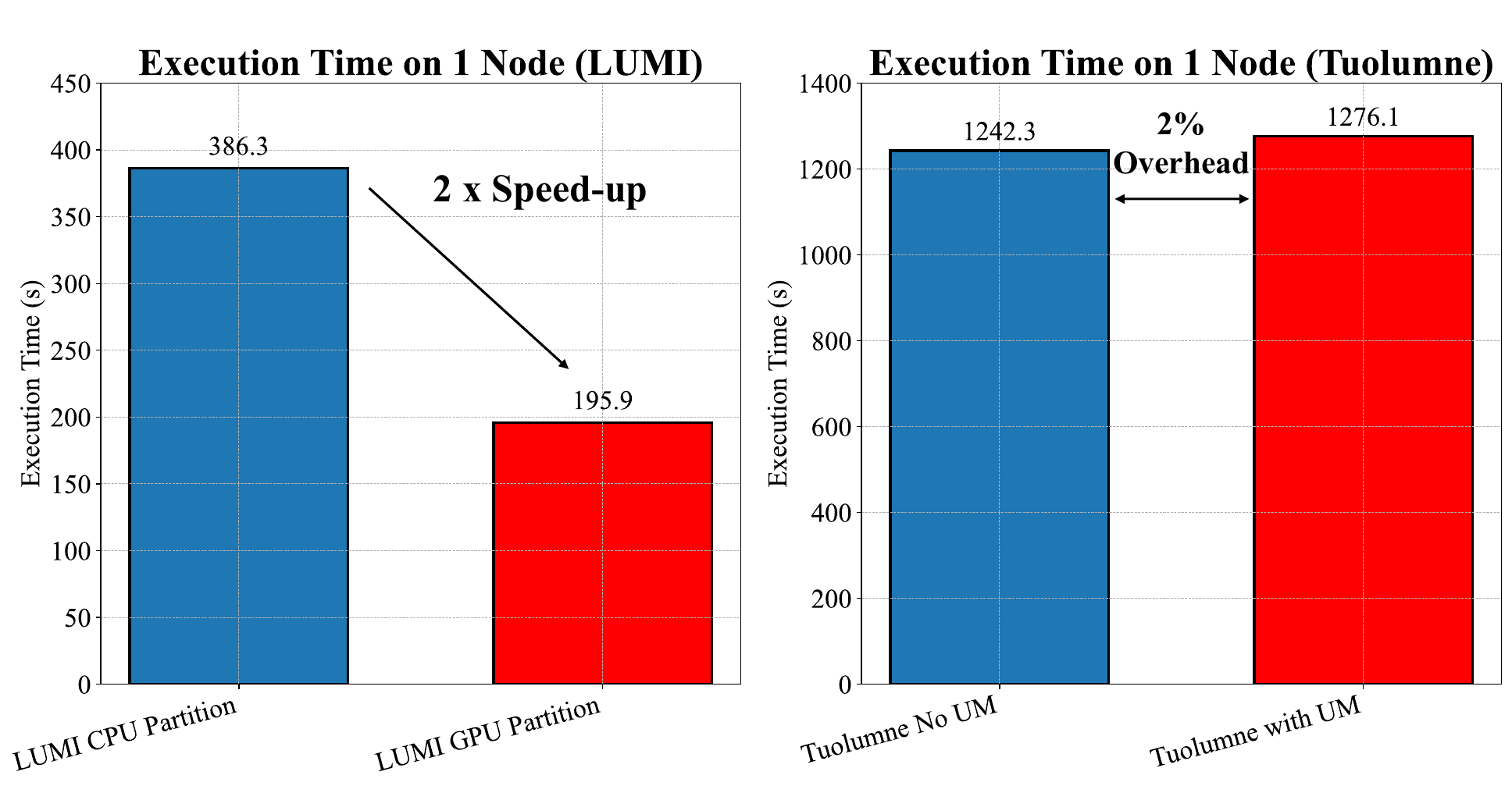}  
    \caption{Left: Execution time of a 500 cycles magnetic reconnection simulation on LUMI-C (blue; four AMD Epyc Trento CPUs) vs. LUMI-G (red; four AMD MI250x GPUs), showing nearly 2 $\times$ improvement. Right: Execution time on one node of Tuolumne with and without unified memory on an AMD MI300A APU. Unified memory introduces negligible overhead while supporting simulations with significantly larger numbers of particles.}
    \label{fig:GPUperf}
\end{figure}

Performance improvements obtained by running accelerators are shown in Figure~\ref{fig:GPUperf}. On the left, the execution time for a magnetic reconnection test case is shown for LUMI-C (four AMD EPYC 7763 CPUs with 64 cores each) and LUMI-G (four AMD MI250x GPUs). A \ul{nearly 2 $\times$ speed-up is achieved on the GPU platform}. On the right, we compare the execution time on Tuolumne with and without unified memory on an AMD MI300A APU, showing negligible computational overhead, which allows for simulation with a larger number of particles.

\subsection{Technical Contribution II: Physics-Aware Compression}
PIC simulations generate vast amounts of data that span from grid-based quantities such as electric and magnetic fields, densities, currents, and pressures, to detailed information for each particle, including positions, velocities, and additional statistical measures. Among these, the particle distribution functions -- crucial for understanding the phase-space dynamics and the coupling between particles and waves -- are of paramount importance. However, since modern simulations can involve hundreds of trillions of particles, saving full particle data at every time step is impractical. Instead, particle information is stored only at a few select intervals to support checkpoint/restart operations and enable the reconstruction of the distribution functions in post-processing analyses. Even in this case, processing full-scale simulation data in the range of petabytes remains a non-trivial task.

In the introduction of this paper, the full phase-space distribution function \(f(\mathbf{x}, \mathbf{v}, t)\) was detailed. Here, we define the velocity-space distribution function at a fixed time \(t_0\) as the spatial integral of the full distribution,
$f(\mathbf{v})~=~\int f(\mathbf{x}, \mathbf{v}, t_0) \,\text{d}\mathbf{x}$, over one simulation domain. For brevity, we denote this marginal distribution simply as \(f(\mathbf{v})\) in the following discussion.

\noindent \textbf{Implemented Solution:} To tackle the challenge of managing such large datasets, we implement the GMM compression scheme~\cite{chen2021unsupervised} executed entirely on the GPUs and APUs to compress the particle data and obtain an analytical expression for the 3D $f(\mathbf{v})$ in each subdomain~\cite{Andong}. Rather than applying the GMM directly to the raw particle data, we first perform a three-dimensional binning operation in velocity space. This binning leads to a histogram representation of the electron and ion distribution functions, reducing the data complexity. In our approach, the velocity-space distribution function \(f(\mathbf{v})\) is approximated as a weighted sum of \(M\) Gaussian components: $f(\mathbf{v}) \approx \sum_{i=1}^M \alpha_i\, \mathcal{N}\!\left(\mathbf{v}\,\middle|\, \boldsymbol{\mu}_i, \boldsymbol{\Sigma}_i\right),$
where \(\alpha_i\) are the mixture weights, \(\boldsymbol{\mu}_i\) denote the mean velocities, and \(\boldsymbol{\Sigma}_i\) represent the covariance matrices, which characterize the thermal spread of the distribution. The parameters \(\{\alpha_i, \boldsymbol{\mu}_i, \boldsymbol{\Sigma}_i\}\) are estimated using the Expectation-Maximization (EM) algorithm~\cite{mclachlant97EM_algorithm} and store the information of the compressed distribution functions.

The choice of the GMM algorithm is justified by fundamental physical considerations. Under most plasma conditions, electron and ion distribution functions exhibit near-Gaussian behavior, and variations in the Gaussian parameters correspond directly to physical phenomena. For instance, an increase (decrease) in covariance matrix values indicates heating (cooling), while non-zero mean velocities may signal the presence of particle beams. Given this direct relationship between mathematical parameters and physical phenomena in the magnetosphere, we refer to our method as \emph{physics-aware} compression. For parallel I/O operations on compressed data, we employ the ADIOS 2 framework~\cite{godoy2020adios}.

\begin{figure}[t]
    \centering
    \includegraphics[width=\linewidth]{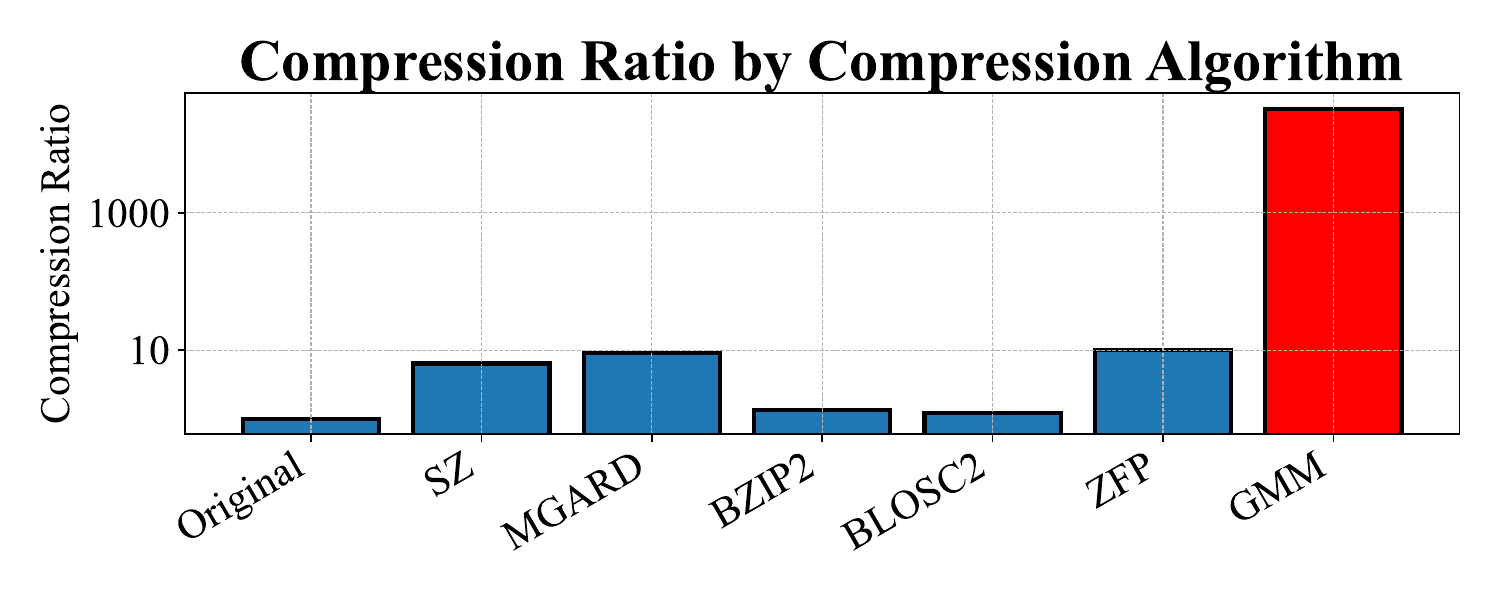}
    \caption{Compression ratios achieved by different compression algorithms. The proposed GMM-based method (red bar) outperforms other established compression methods in terms of compression rate in the specific task of compressing particle distribution functions.}
    \label{fig:compression}
\end{figure}

\noindent \textbf{Key Results and Impact:} The experimental results, summarized in Figure~\ref{fig:compression}, show that our physics-aware compression method based on GMM, tailored for storing distribution functions, outperforms several established compression schemes. In particular, \ul{our approach achieves compression ratios above 1,000 with minimal accuracy loss}. Alternative general-purpose floating-point compression methods such as SZ, MGARD, BZIP2, BLOSC2, and ZFP~\cite{li2018data,di2024survey} lead to compression ratios closer to 10. This large GMM compression ratio is achieved while maintaining a minimal loss of accuracy, confirmed by Jensen–Shannon divergence values of $O(10^{-2})$.
The full GPU implementation of our GMM-based compression enables a rapid in-situ data compression, which overlaps with the Maxwell solver and thus minimally affects execution time (10\% overhead). Thanks also to the 3D pre-binning of the velocity-space, our physics-aware compression scheme has a competitive computational cost compared to other compressors in ADIOS-2~\cite{godoy2020adios}. We note that the compression efficiency and cost of all methods, including our GMM approach, depend on configuration parameters that can be tuned to balance the trade-off between compression ratio, accuracy, and execution time according to the specific needs of the simulation.

\subsection{Technical Contribution III: Integrated Data Analysis}
Magnetosphere simulations produce large volumes of data, making data analysis a challenging task. Thus, identification of the spatial regions and time intervals where key physical phenomena, such as particle heating, acceleration, and anisotropy, occur is essential. These events cannot be directly inferred from field or moment values. However, the high-cadence storage of electron and ion distribution functions enables automatic data analysis and anomaly detection.

\noindent \textbf{Implemented Solution:} We leverage compressed electron and ion distribution functions for in-transit data analysis~\cite{kelling2025artificial}. A dedicated Python script monitors the arrival of new compressed data. The script reconstructs the distribution functions and computes a set of statistical descriptors: mean, variance, skewness, kurtosis, differential entropy, and Kullback–Leibler (KL) divergence between two successive distribution functions.
In particular, we employ anisotropy and differential entropy to identify the onset of particle acceleration and heating. These metrics are extrapolated from a velocity distribution function $ f(\mathbf{v})$ as follows. We define the probability density function (pdf) as $p(\mathbf{v}) = f(\mathbf{v})/ \int f(\mathbf{v}) \, d\mathbf{v}$, and from the variance of the pdf, we calculate the anisotropy index. The differential entropy, $h$, is then computed as $h(p) = -\int p(\mathbf{v}) \log p(\mathbf{v}) \, d\mathbf{v}$. Temporal trends and change points in these metrics are analyzed using the \texttt{ruptures} module for change point detection~\cite{truong2018ruptures}.

\begin{figure}[t]
    \centering
    \includegraphics[width=\linewidth]{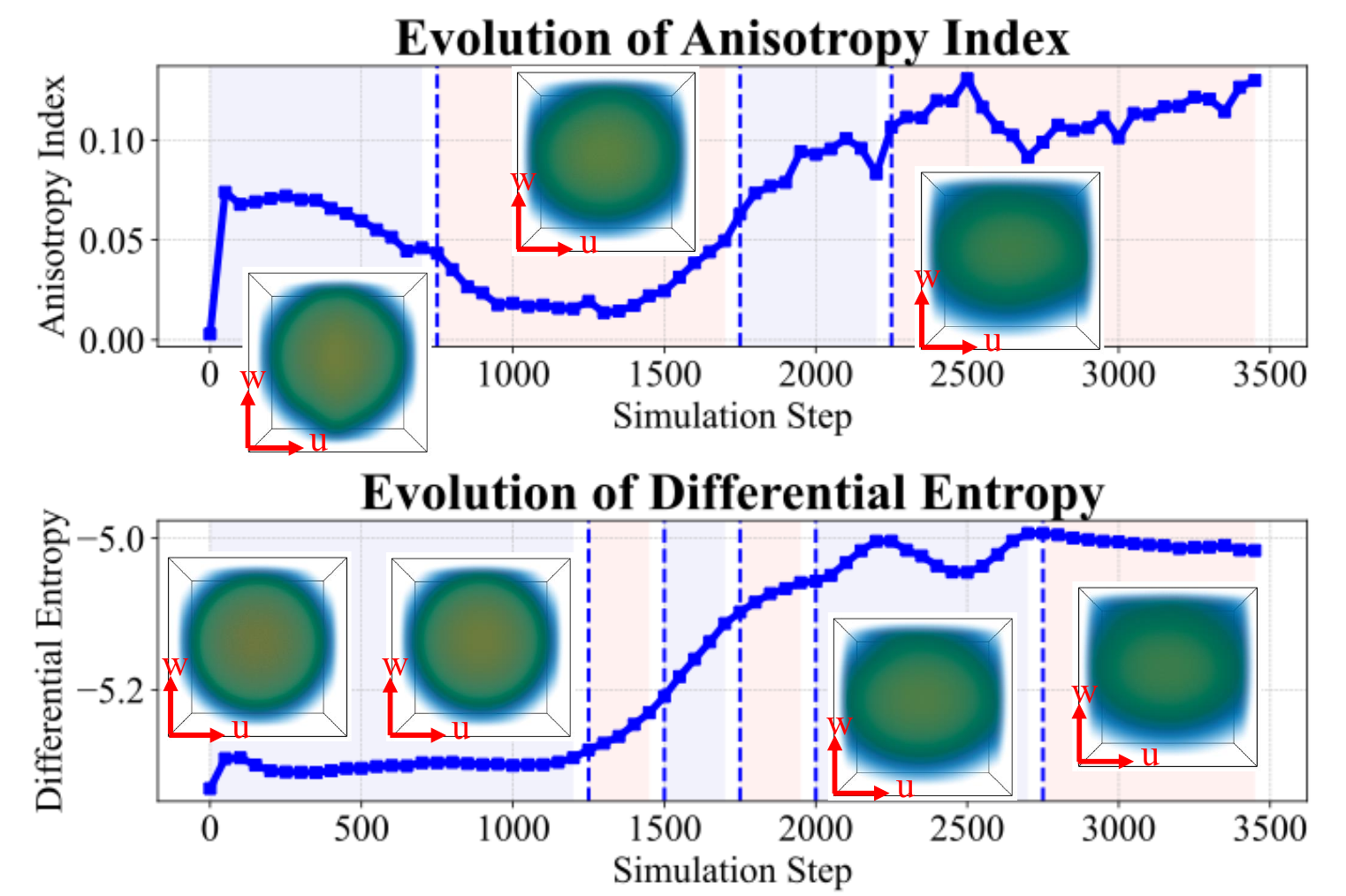}  
    \caption{Detected change points (dashed lines) in the evolution of variance anisotropy and differential entropy of the electron pdf in the proximity of the diffusion region. 3D electron distributions are superimposed in connection with the change points. The electron diffusion region exhibits a 3D distribution function elongated in the $u$ direction (velocity component in the x direction), characteristic of electron beams exiting the magnetic reconnection line.}
    \label{fig:ruptures}
\end{figure}

\noindent \textbf{Key Results and Impact:}  The usage of change point algorithms successfully tracks dynamic transitions in the distribution functions. In particular, statistical indicators, such as differential entropy and anisotropy between variances, provide clear signals of physical changes. Figure~\ref{fig:ruptures} illustrates the application of this approach to study the electron distribution function in the electron diffusion region during magnetic reconnection. The detected change points mark the onset of significant variations in the electron distribution functions. This integrated analysis can be performed in real-time or in post-processing, improving our ability to pinpoint and study dynamic processes in the magnetosphere.

\section{Experimental Setup}
We assess the weak and strong scaling of our \texttt{{\small iPIC3D}} simulations through performance metrics, including time-to-solution and scaling efficiency. All reported measurements reflect the full application performance, inclusive of initialization, MPI communication, computational kernels, and I/O.
For the scaling tests, we utilize a representative 3D simulation of magnetic reconnection, inspired by the GEM challenge~\cite{birn2001geospace}, which is extended to a 3D domain that mimics Earth's magnetotail. This scenario serves as a \textit{de facto} standard benchmark for kinetic simulations in space physics.

In both weak and strong scaling studies, simulations are executed at least three times, except for the largest run on 32,768 APUs, executed only once. The recorded execution times include the overhead of reading input files and MPI initialization. In the weak scaling tests, the simulation domain's size is increased proportionally to the number of computational nodes, maintaining a constant number of particles and computational workload per node. For strong scaling, we choose a fixed simulation problem size that fully occupies GPU memory on a small number of nodes. We then progressively increase the number of nodes while keeping the overall simulation domain and particle count constant.

To compute the sustained performance in FLOP/s on El Capitan, we follow a methodology used in previous work~\cite{fedeli2022pushing}: we use profiling data via the AMD \texttt{rocprof} tool from a single El Capitan node to determine the total floating-point operations in double precision performed in each computational cycle. We scale this value by the number of nodes used in the largest simulation, including the measured parallel efficiency, to account for deviations from ideal scaling. For peak performance measurement, we focus on the particle mover routine -- the most computationally intensive part. We analytically count the number of floating-point operations per particle mover and divide this number by its execution time, measured with timers.

Finally, we evaluate I/O performance by measuring checkpointing efficiency. Checkpointing times were recorded separately from computational kernels to identify bottlenecks and to optimize.

\section{Performance Results}
In this section, we present the performance results of \texttt{{\small iPIC3D}} simulations on several leading supercomputers, with focus on the El Capitan system. Table~\ref{tab:supercomputer_specs} summarizes the key system characteristics for the supercomputers utilized in our performance tests, highlighting El Capitan for clarity.

\begin{table}[t]
\centering
\caption{Key system characteristics of the supercomputers used in performance tests.}
\renewcommand{\arraystretch}{1.3}
\begin{adjustbox}{max width=\columnwidth,center}
\large
\begin{tabular}{lccc}
\hline
Supercomputer & Top500 \# & Computational Nodes & Network Model \\
\hline
\rowcolor{gray!20}
El Capitan     & 1   & 4 $\times$ AMD MI300A   & 	Slingshot-11 \\
\hline
LUMI-G           & 8   & 4 $\times$ AMD MI250x + 1 $\times$ AMD EPYC        & Slingshot-11 \\
Leonardo Booster       & 9   & 4 $\times$ NV. A100 + 1 $\times$  Intel Xeon  & HDR100 Infiniband\\
Tuolumne       & 10  & 4 $\times$ AMD MI300A           & Slingshot-11 \\
MareNostrum 5 ACC& 11  & 4 $\times$ NV. H100  + 2 $\times$ Intel Sapphire Rapids  & Infiniband NDR \\
Lassen         & 72  & 4 $\times$ NV. V100    +  1 $\times$IBM Power9             & EDR Infiniband \\
\hline
\end{tabular}
\end{adjustbox}
\label{tab:supercomputer_specs}
\end{table}

The primary performance measurements are carried out on El Capitan, \ul{utilizing up to 32,768 AMD MI300A APUs, corresponding to a total of 8,257,536 cores}. Additionally, we perform scaling tests on smaller systems, including LUMI, Leonardo, Tuolumne, MareNostrum 5, and Lassen. These tests involve up to 1,000 -- 2,000 GPUs/GCDs/APUs, and allow us to evaluate the performance across various supercomputer node architectures and network configurations. Detailed information about the hardware, software and compilation environment is provided in the paper data artifacts~\footnote{\href{https://github.com/iPIC3D/iPIC3D-GPU/blob/scaling/scalingConfigurations.md}{iPIC3D scaling artifact repository}}.

\subsection{Strong \& Weak Scaling}
\begin{figure}[t]
    \centering
    \includegraphics[width=\linewidth]{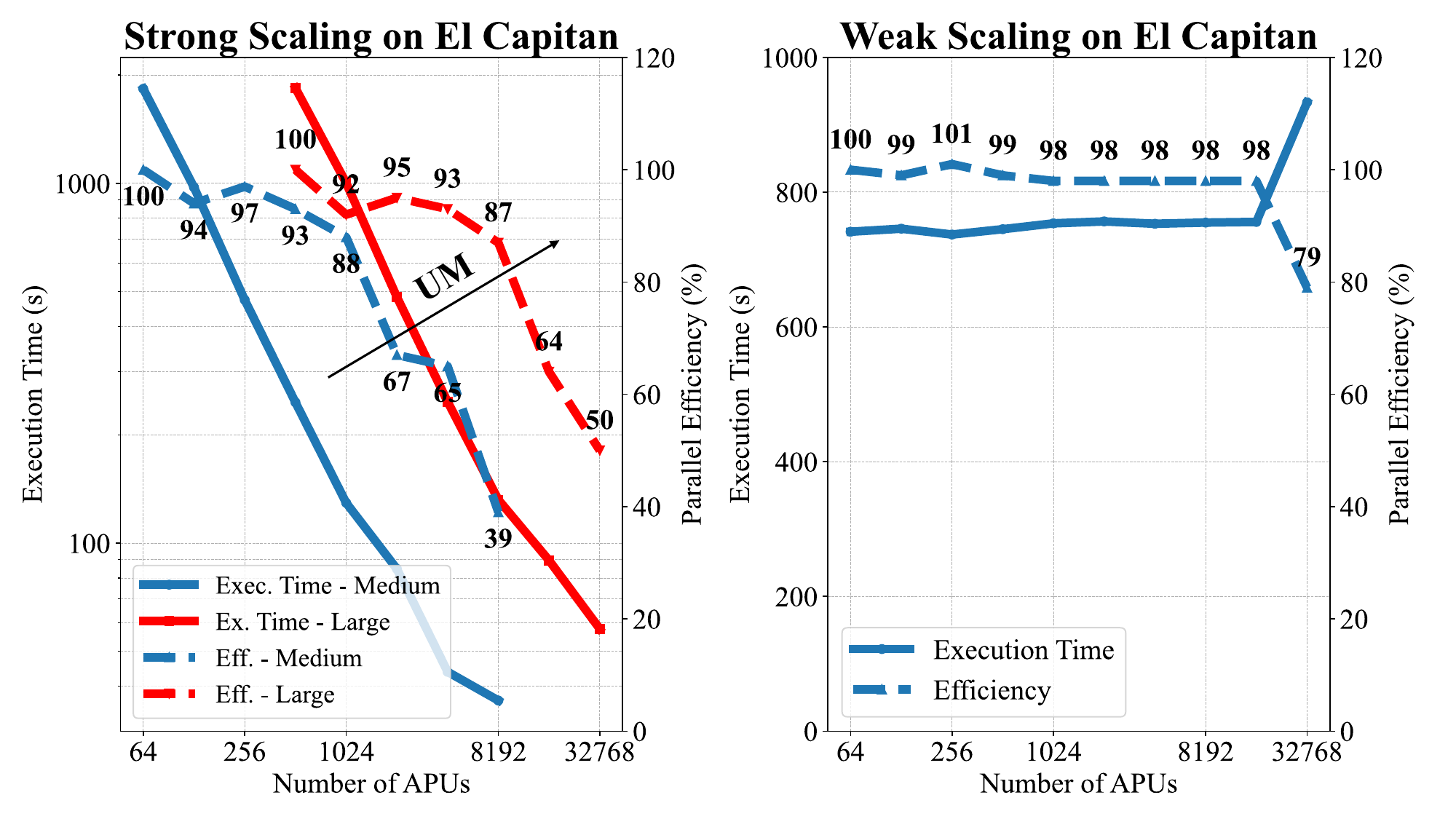}
    \caption{Scaling performance on El Capitan. Left Panel (Strong Scaling): blue continuous and dashed lines denote execution time and parallel efficiency, respectively, for the medium simulation. Red lines represent the large simulation. Right Panel (Weak Scaling): Execution time remains nearly constant as the number of APUs increases, and parallel efficiency remains close to 100\% for small to moderate scales, dropping to approximately 79\% at 32,768 APUs.}
    \label{fig:scaling}
\end{figure}
\begin{table}[t]
\centering
\caption{Sustained and peak performance achieved in production-scale runs.}
\renewcommand{\arraystretch}{1.3}
\begin{adjustbox}{max width=\columnwidth,center}
\large
\begin{tabular}{ccc}
\hline
Nodes Used & Projected Peak Performance  & Projected Sustained Performance  \\
\hline
\rowcolor{gray!20}
 32,768 & 9.64 $\times 10^{17}$ (964 PFLOP/s) & 2.24 $\times  10^{16}$ (22.4 PFLOP/S) \\
\hline
\end{tabular}
\end{adjustbox}
\label{tab:flops_results}
\end{table}
We evaluate the computational efficiency of our implementation for strong and weak scaling tests on El Capitan.

\noindent \textbf{Strong Scaling:} In the strong scaling experiments, the total problem size is fixed while the number of APUs is increased. Two simulation configurations are compared:
\begin{itemize}
    \item \textbf{Medium Simulation:} This configuration uses all available memory by employing separate CPU and GPU memories. As a result, additional buffers and data copies are required.
    \item \textbf{Large Simulation:} This configuration uses a unified APU memory, eliminating extra memory copies and buffer overhead. In particular, using unified memory, we can perform a simulation with 1.8 $\times$ more computational particles than the simulation treating the CPU and GPU as having separate memories. 
\end{itemize}
The plot in Figure~\ref{fig:scaling} (left panel) shows that, for the medium simulation, the execution time decreases substantially as the number of APUs increases, with parallel efficiency remaining high (approximately 94\% to 97\% at lower scales) before gradually declining (to approximately 39\% at the highest APU counts). In contrast, while the large simulation exhibits higher absolute execution times (on the order of 1,000 seconds at the baseline), it benefits from the unified memory design, which allows scaling to larger systems. However, its parallel efficiency starts lower (approximately 67\%–65\%) and drops further at the highest scales (down to approximately 59\%), likely due to increased communication overheads at extreme parallelism.

\noindent \textbf{Weak Scaling:} In the weak scaling tests, the problem size is increased proportionally with the number of APUs so that each processing element maintains a constant workload. As shown in Figure~\ref{fig:scaling} (right panel), the execution time remains nearly constant across the tested range, which is indicative of good weak scaling behavior. The parallel efficiency is close to 100\% for lower APU counts and only gradually decreases at 32,000 devices, reaching around 79\%. This slight degradation at scale is attributed to the impact of other jobs interfering with a simulation that uses almost all system nodes.



We report the projected sustained and peak double-precision performance achieved during the runs on El Capitan using MI300A APUs. Sustained performance was obtained using profiler-based FLOP counts per node and scaled. The results are summarized in Table~\ref{tab:flops_results}.

\begin{figure}[t]
    \centering
    \includegraphics[width=\linewidth]{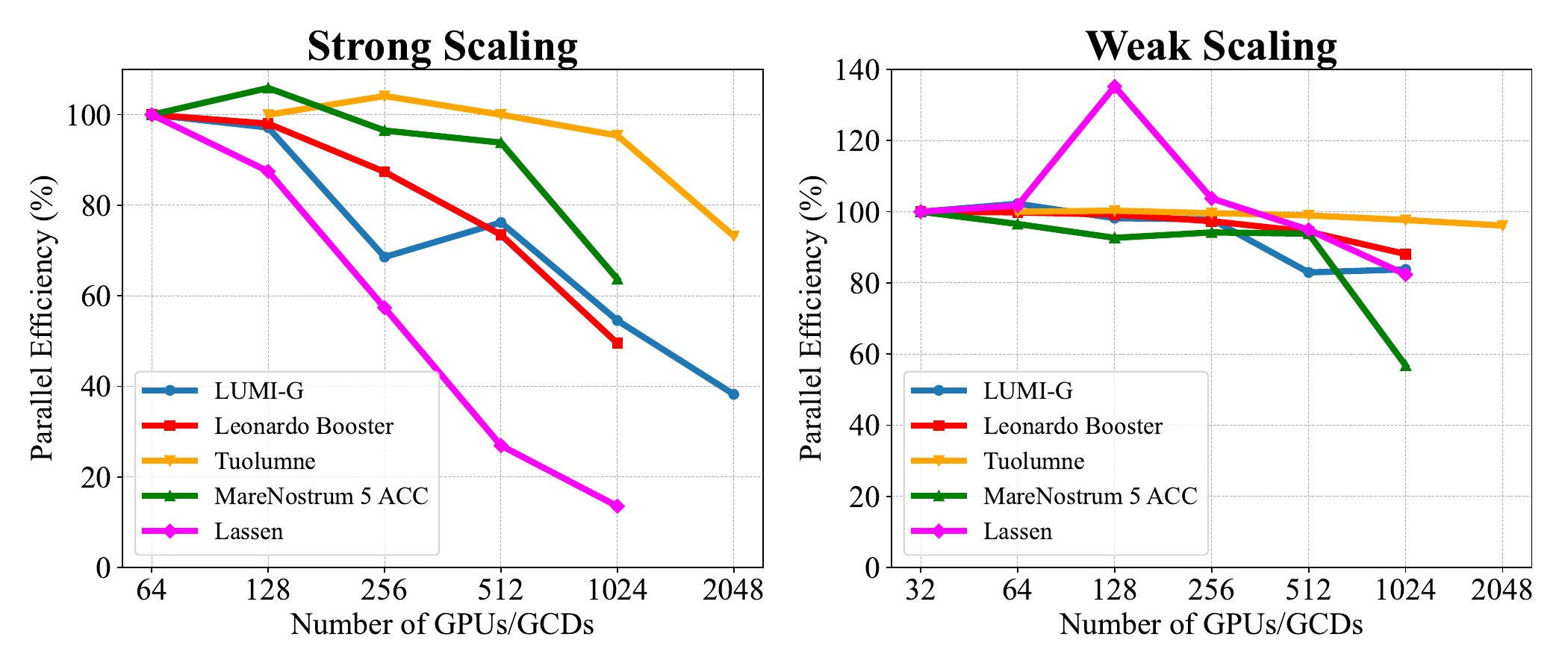}
    \caption{Parallel efficiency for strong and weak scaling on LUMI-G, Leonardo Booster, Tuolumne, MareNostrum 5 ACC, and Lassen supercomputers. Left panel: Strong scaling efficiency from 64 to 2,048 GPUs/GCDs. Right panel: Weak scaling efficiency for 32 to 2,048 GPUs/GCDs.}
    \label{fig:scaling_efficiency}
\end{figure}

In addition to the large-scale tests on El Capitan, we conduct performance experiments on five additional supercomputing systems: LUMI, Leonardo, Tuolumne, MareNostrum 5 ACC, and Lassen. Figure~\ref{fig:scaling_efficiency} summarizes the strong and weak scaling efficiencies.  The left panel of Figure~\ref{fig:scaling_efficiency} shows strong scaling efficiency as the number of APUs/GPUs/GCDs increases from 64 to 2,048. High parallel efficiency is maintained at lower node counts on all systems, with a gradual decline at larger scales due to communication overhead. The right panel presents the weak scaling results, where the problem size is increased proportionally with the number of APUs/GPUs/GCDs. Across all systems, the parallel efficiency remains nearly constant, indicating good scalability. For the Lassen supercomputer, an improved network performance with 128 GPUs results in a parallel efficiency 30\% higher than the ideal efficiency.

\begin{figure}[h]
    \centering
    \includegraphics[width=\linewidth]{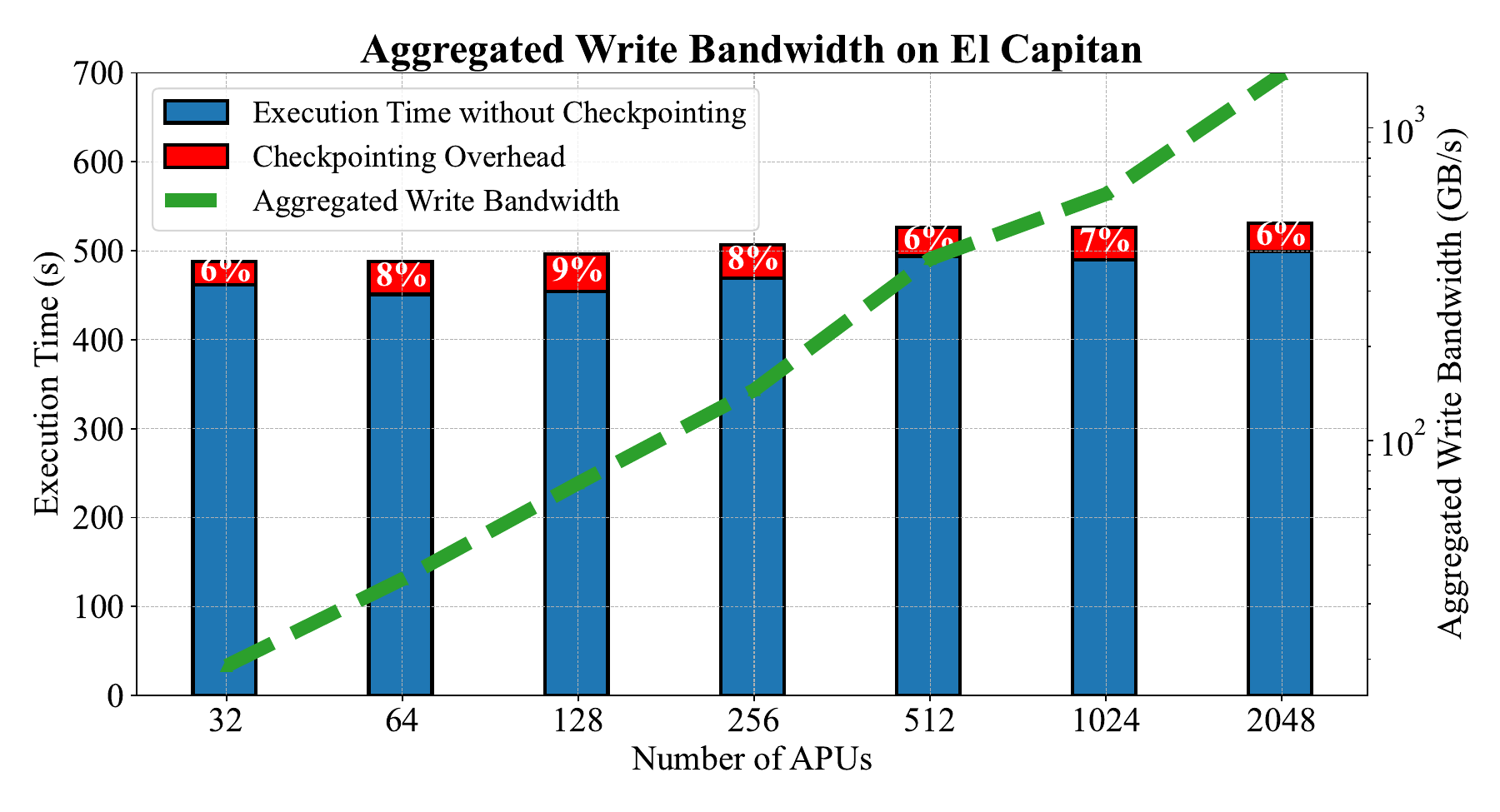}
    \caption{I/O performance and checkpointing overhead on El Capitan. The x-axis shows the number of APUs, the left y-axis indicates the execution time (in seconds) for 500 \texttt{{\small iPIC3D}} cycles of a magnetic reconnection simulation (excluding checkpoint I/O), and the right y-axis shows the aggregated write bandwidth (in GB/s). The red segments on the bars are the additional time required for checkpointing, with overheads annotated between 6\% and 9\%.}
    \label{fig:IO_scaling}
\end{figure}

Efficient I/O and data checkpointing are critical in large-scale PIC simulations. In addition, we analyze the I/O performance and checkpointing overhead measured during a weak scaling run of 500 \texttt{{\small iPIC3D}} cycles of magnetic reconnection. The results are presented in Figure~\ref{fig:IO_scaling}. The x-axis indicates the number of APUs used, ranging from 32 to 2,048. The left y-axis shows the execution time (in seconds) for the simulation cycles, while the right y-axis represents the aggregated write bandwidth in GB/s. Each bar in the plot represents the execution time for 500 cycles without the I/O cost for checkpointing particles and fields. The red segments in the bars indicate the additional time incurred by checkpointing. The checkpointing overhead is annotated on each bar with percentages ranging from 6\% to 9\%. We find that as the number of APUs increases, the execution time remains approximately constant, demonstrating effective weak scaling with I/O. Simultaneously, the aggregated write bandwidth increases, showing higher data throughput as the system size grows. Importantly, the checkpointing overhead remains relatively low across all scales. Even at the largest system sizes, the additional cost due to checkpointing is relatively small (approximately 6\% to 9\%). 

\begin{figure*}[t]
    \centering
    \includegraphics[width=0.7\linewidth]{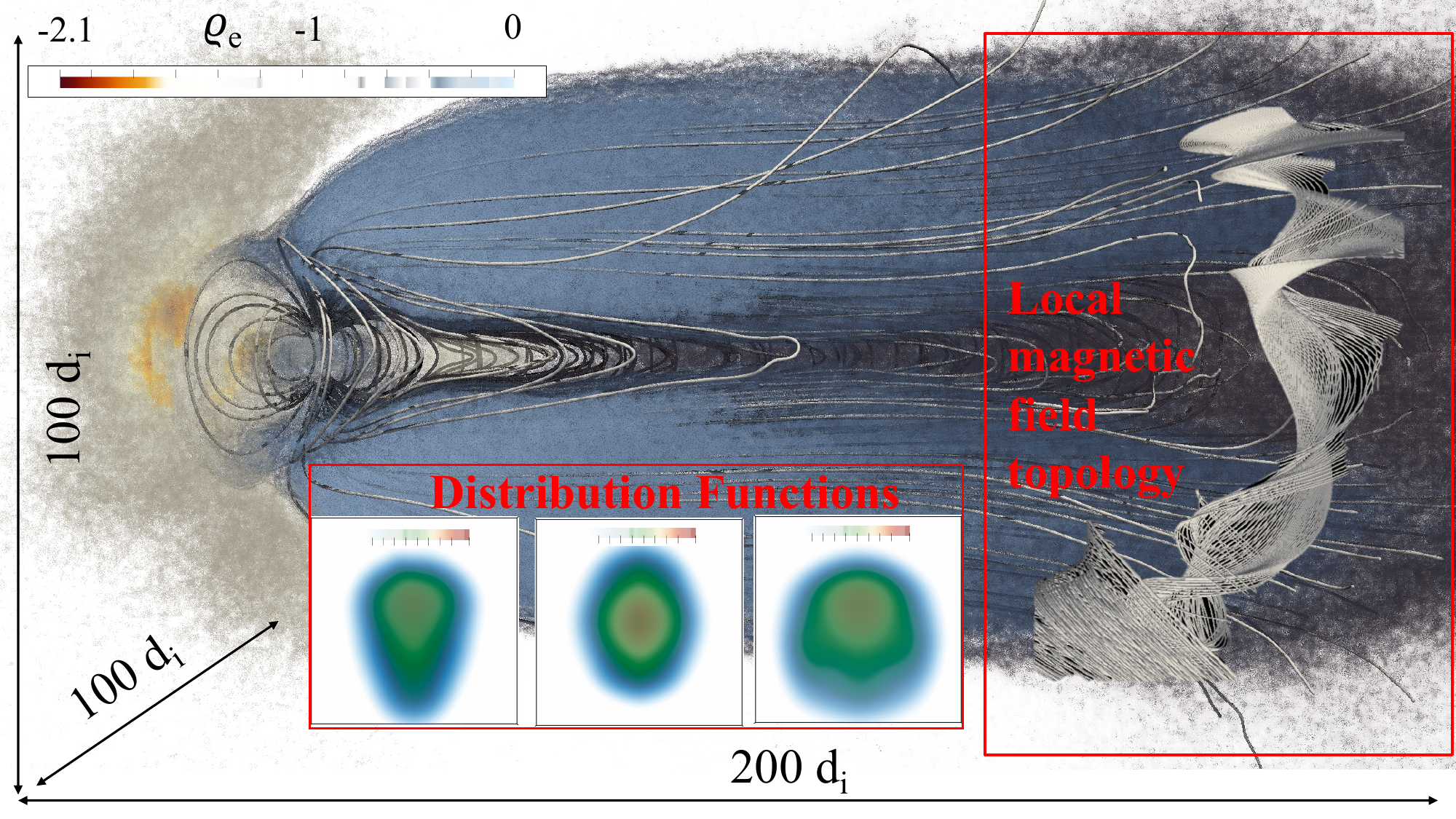}  
    \caption{Volume rendering of the electron density $\rho_e$ in a large-scale magnetosphere of size $200\,d_i \times 100\,d_i \times 100\,d_i$, with superimposed magnetic field lines. The simulation uses 320 $\times$ 160 $\times$ 160 cells and initially one billion particles. The insets show a zoomed-in view of the local magnetic field topology of a flux rope in the magnetotail and ion distribution functions observed in the magnetotail reconnection~\cite{markidis2013kinetic,markidis2012collisionless}. The global magnetospheric simulations capture complex three-dimensional topologies that include reconnection sites and bow shocks, resolve microscales, providing distribution functions, solving the micro-macro challenge in magnetosphere simulations.}
    \label{fig:full_magnetosphere}
\end{figure*}

\section{Discussion \& Conclusion}
 
\subsection{Magnetospheric Physics Impact}
Understanding the multi-scale coupling between microscopic electron-scale physics and macroscopic global magnetospheric dynamics remains one of the central challenges in space physics. Unlike traditional approaches that rely on unrealistic parameters to relax numerical stability constraints, our implicit PIC framework enables global kinetic simulations using physically realistic solar wind and plasma parameters. The numerical advances enable us to simulate global magnetospheres with dimensions of several hundred ion skin depths on thousands of APUs, allowing for the modeling of small- to medium-scale planetary magnetospheres, such as those of Ganymede and Mercury, in a physically consistent, fully kinetic regime. An example of such simulations, spanning in space for more than a hundred $d_i$, is shown in Figure~\ref{fig:full_magnetosphere}.

These capabilities establish a new modeling regime for magnetospheric systems. By resolving both global structures and localized kinetic processes, such as reconnection and particle acceleration, the simulations enable the investigation of phenomena that are traditionally examined in isolation. For example, magnetic reconnection is often studied using idealized configurations, such as the Harris current sheet. In contrast, our global simulations provide self-consistent and physically realistic magnetic field topologies and equilibria, enabling the study of reconnection and bow shocks within nontrivial, three-dimensional configurations.

\subsection{HPC and Scientific Grand Challenges}
The fully kinetic simulation of global planetary magnetospheres with realistic parameters is an archetypal exascale challenge. It requires innovations in algorithms beyond standard explicit schemes, performance optimization on distributed heterogeneous supercomputers, and data management and analysis at extreme scales. This paper demonstrates the feasibility of such simulations at scale, exploiting both implicit methods and modern heterogeneous architectures.

From an algorithmic standpoint, the use of implicit PIC methods represents a major advancement over conventional explicit techniques. Implicit discretizations enable significantly larger time steps and grid spacings, eliminating the need to resolve electron Debye lengths and plasma frequencies without compromising accuracy. Although this increases per-step computational complexity -- due to iterative solvers and nonlinear coupling -- overall efficiency is largely improved, by more than three orders of magnitude. Such advances suggest that simulations of Earth's magnetosphere at realistic scales may soon become feasible. However, achieving this goal will require further algorithmic developments, including the integration of additional physics, such as ionospheric coupling and collisional models. 

We also showed that implicit PIC algorithms can be effectively ported to GPU-accelerated and APU-based architectures. Our \texttt{{\small iPIC3D}} code on AMD MI300As achieves a $2\times$ performance improvement over CPU-only configurations. The AMD APU's unified memory architecture enables an increase in the number of computational particles that can be processed at a negligible computational cost. 

Data movement and analysis at exascale pose additional challenges.  To address this, we developed a physics-aware, in-situ lossy compression scheme based on GMM, specifically designed for particle velocity distribution functions. This method achieves compression ratios exceeding 1,000$\times$ while preserving key physical features. The compression is integrated into the simulation workflow, supporting real-time anomaly detection and change-point analysis.

Finally, our results highlight the importance of moving beyond monolithic application models. Fully leveraging modern supercomputers and accelerators requires workflows that integrate simulation, data compression, and in-situ analysis. This comprehensive approach enhances system utilization by ensuring the effective use of GPU and CPU resources in each phase of the application.

\section*{Acknowledgments}
{\small This work was performed under the auspices of the U.S. Department of Energy by Lawrence Livermore National Laboratory under Contract DE-AC52-07NA27344 and was supported by the LLNL-LDRD Program under Project No. 24-SI-005 (LLNL-CONF-2004570). We acknowledge the EuroHPC Joint Undertaking for awarding this project access to the EuroHPC supercomputer LUMI, hosted by CSC (Finland) and the LUMI consortium, Leonardo, hosted by CINECA (Italy), MareNostrum 5, hosted by BSC (Spain) through a EuroHPC Regular Access call.}
\bibliographystyle{ieeetr}
\bibliography{GBiPIC3D}

\end{document}